# Modeling and Predicting Students' Engagement Behaviors using Mixture Markov Models

**Rabia Maqsood · Paolo Ceravolo · Cristóbal Romero · Sebastián Ventura**

**Abstract** Students' engagement reflect their level of involvement in an ongoing learning process which can be estimated through their interactions with a computer-based learning or assessment system. A pre-requirement for stimulating student engagement lie in the capability to have an approximate representation model for comprehending students' varied (dis)engagement behaviors. In this paper, we utilized model-based clustering for this purpose which generates $K$ mixture Markov models to group students' traces containing their (dis)engagement behavioral patterns. To prevent the Expectation-Maximization (EM) algorithm from getting stuck in a *local maxima*, we also introduced a new initialization method named as K-EM. The proposed method initializes the EM algorithm using the results of a preliminary K-means clustering algorithm performed on students' logged problem-solving actions. We performed an experimental work on two real datasets using the three variants of the EM algorithm: the original EM, emEM, K-EM; and, non-mixture baseline models for both datasets. The proposed K-EM method has shown very promising results and achieved significant performance difference in comparison to the other approaches particularly using the Dataset1 (which contains small length traces in contrast to the Dataset2). Hence, we suggest to perform

R. Maqsood (  )
Department of Computer Science, National University of Computer and Emerging Sciences, Pakistan
E-mail: rabia.maqsood@nu.edu.pk

P. Ceravolo
Computer Science Department, Universitá degli Studi di Milano, Milan, Italy

C. Romero
Department of Computer Science and Numerical Analysis, University of Cordoba, Cordoba, Spain

S. Ventura
Department of Computer Science and Numerical Analysis, University of Cordoba, Cordoba, Spain

further experiments using large dataset(s) to validate our method. Additionally, visualization of the resultant $K$ mixtures (or clusters) through first-order Markov chains reveal very useful insights about (dis)engagement behaviors depicted by the students. We conclude the paper with a discussion on the usefulness of our approach, limitations and potential extensions of this work.

## 1 Introduction

Although there are a lot of definitions available to date in the literature for students' engagement, it is generally referred as *active participation* in an ongoing task or process. In other words, engagement reflects a student's level of involvement in a learning process. It is thus a crucial notion that becomes even more important when students are interacting with a computer-based learning or assessment system as the main objective of these systems is to facilitate students in learning and improving their learning outcomes. However, if a student does not show interest or engage appropriately during a learning process, he/she may observe failure or degradation in performance (Cocea and Weibelzahl, 2007) and consequently abandon the learning process. There is evidence to show that students' *online engagement* (which is estimated through their behaviors while interacting with a learning/assessment environment) is positively correlated with good performance scores in standardized exams (Pardos et al., 2014) and students' academic outcomes (Vogt, 2016). Recently, researchers have been showing great interest in measuring student online engagement after realizing that the student's knowledge gap cannot be addressed easily if he/she does not show interest while interacting with a learning environment (Desmarais and Baker, 2012).

Computer-based learning enables tracking students' activities at the micro-level, i.e. event by event, but this information can be exploited only if events are encoded with a suitable representation model. In this work, we aim to model, analyze and predict students' (dis)engagement behaviors in confidence-based assessment, which requires students to specify their confidence level with each submitted answer (Gardner-Medwin and Gahan, 2003). Thus, confidence-based assessment provides two outcome measures for student evaluation, that is, a student's submitted response to a question (which can be either correct or incorrect) and his/her associated confidence level in that response (e.g., as high or low). Maqsood and Ceravolo (2018) discussed the importance and usefulness of this two-dimensional assessment paradigm and highlight the need for capturing students' dynamic behaviors during assessment. A mapping between students' (dis)engagement behaviors and the two performance measures of confidence-based assessment (i.e., a student's response correctness and associated confidence level), is presented in (Maqsood et al., 2019) based on theoretical reasoning. The proposed classification scheme defined six (dis)engagement behavioral patterns, each representing a student's positive or negative engagement during assessment. Specifically, the six behavioral patterns namely: High Knowledge (HK), Less Knowledge (LK), Fill-knowledge Gap (FG), Knowledge Gap (KG), Learn (LE), and, Not Interested (NI); are defined using the following three problem-solving actions: a student's response correctness (correct or incorrect), confidence level (high or low) specified for each submitted answer, and, a followed feedback-seeking activity (whether a student requested task-level feedback or not for an answered question), see Section 2.1 for the details. We refer to these categories as "(dis)engagement behavioral patterns" (sometimes simply referred as behavioral patterns in this paper) as they do not represent sole action a student performs, but instead, each discrete label is a composite of three attributes of a student's problem-solving behavior and thus reflecting his/her (dis)engagement behavior in the on-going assessment process.

In this work, our first objective is to construct a mechanism to model students' engagement/disengagement behaviors that can be used to analyze their sequential problem-solving traces, wherein each activity is represented by a behavioral pattern belonging to the set $P$, where $P = \{$HK, LK, FG, LE, KG, NI$\}$. For this purpose, we utilized model-based clustering to group multivariate categorical time series data representing students' traces, each containing behavioral patterns of different lengths. The Expectation-Maximization (EM) algorithm (Dempster et al., 1977) used for constructing mixture Markov models struggle for finding *global* maxima and hence the initialization method can play an important role in finding a best solution (Michael and Melnykov, 2016; Hu, 2015). Therefore, we have performed an experimental work on two different variants of the EM algorithm along with our proposed initialization method, called K-EM. All the methods are run on two real datasets taken from two different studies conducted with undergraduate students using computer-based assessment systems.

The second objective of this research work is to predict students' future (or next) behavioral patterns so that students with varying needs can be identified and referred for appropriate intervention (by a teacher or an adaptive system, if needed). For example, KG and NI patterns respectively show disengaged behaviors of high and low confident students requiring different course of actions to re-engage them in the assessment process. We have reported the details of our experimental work and the achieved results in this paper. Our proposed initialization method, K-EM has shown promising results; however, further experiments on large dataset(s) are required for validation. Finally, visualization of the resultant mixtures (i.e., Markov models) provide thoughtful insights for theoretical and practical reasoning as discussed in detail in the paper. We also suggest that the resultant Markov chains can be easily interpreted by class teachers if the proposed methodology is implemented in a computer-based assessment tool in the future. That will allow a class teacher to get timely feedback about his/her students after each assessment session (e.g., a computer-based quiz). Additionally, our proposed methodology can be extended further to create students' personalized behavioral profiles, as discussed in Section 9.

The remainder of this paper is structured in the following manner. A brief background on student (dis)engagement and model-based clustering is given in Section 2. Our proposed initialization method for EM algorithm is described in Section 3. Section 4 contains description of the two datasets, details of our methodology for the experimental work and predictive models. The experimental setup and evaluation metrics for comparing different algorithms are explained in Section 5. Subsequently, in Section 6, we present the detailed results. Finally, we have shown the resultant mixture Markov models with their interpretation representing the students' (dis)engagement behaviors during assessment in Section 7. Related works on determining student (dis)engagement and probabilistic approaches used for similar problems are discussed in Section 8. We conclude the paper in Section 9 with a summary, limitations, usability and potential extensions of this work.

## 2 Background

The concept of *school engagement* started getting attraction in the late 90's through the realization of the existence of some factors that might have played a role in students' poor academic performance and high rate of dropouts (Fredricks et al., 2004). Likewise, earlier works are primarily based on theoretical reasoning with a focus on developing theoretical models and frameworks that may be useful to build a connection between students' actions and their thought (or cognitive) process. And, the identified relation(s) can be helpful to understand the reasoning behind different actions performed by a student. Fredricks et al. (2004) described engagement as a multifaceted construct, comprising the following three types: *cognitive engagement* refers to the investment of effort and thoughtfulness to comprehend complex learning ideas and concepts; *emotional engagement* focuses on the student's positive and negative reactions to the environment; and, *behavioral engagement* draws on the idea of students' participation in learning activities. These three dimensions of engagement are well-accepted and widely studied in the literature. Online engagement is also referred as "behavioral engagement" in the literature which relates to a student's participation in a learning environment and is estimated through his/her actions (Anderson, 2017).

Besides theoretical frameworks (for cognition), in large it is now recognized that students' actions with a computer-based learning environment also reflect their "engagement" in an ongoing learning process (Beal et al., 2007). Bouvier et al. (2014) proposed a quantitative approach to analyze and monitor engagement behaviors using a trace-based method that exploits users' logged interactions with interactive systems. The idea is to transform low-level raw traces into useful high-level abstractions of different engagement behaviors. Beal et al. (2006) adopted the notion of students' active participation in a current task for defining engagement. They determined student engagement from a set of three student-system interactions, namely: *response correctness*, *time spent per problem* and *help usage*. Students' problem-solving actions were classified into five different engagement levels including: *Independent-a*, *Independent-b*, *Guessing*, *Help abuse* and *Learning*. Another self-defined classification scheme for categorizing students logged activities into engaged and disengaged behaviors is proposed by (Brown and Howard, 2014). They relied on data analyzes to define two engagement classes, referred to as, *on-task* and *off-task*. Pardos et al. (2014) studied students' behavioral engagement along with affective states using students logged interactions with a mathematics tutoring system (called ASSISTments). The automated behavioral detector model aims to identify two specific behavioral events depicting students' active (or in-active) participation during assessment, namely: *off-task* and *gaming* behaviors. In Hershkovitz and Nachmias (2009), students' logged data was collected from an online vocabulary tool, which was analyzed visually by human experts to identify the important variables relating to their theoretical framework of motivation. Then, different variables were grouped by similarity using the *Hierarchical* clustering algorithm. The cluster group containing *time on task* (percentage) and *average session duration* variables were mapped to students' engagement behaviors. Besides engagement behavior detection, there are several other works on analyzing students' logged data to gain better understating about their usage of learning environments. However, we have specifically reviewed the ones which targeted student engagement and/or behavior detection.

In this work, our focus is on determining and analyzing students' engagement behaviors through their logged data captured by a computer-based assessment system. It is one of the most popular methods for data collection in the educational domain to analyze students' problem-solving actions due to its uninterrupted nature, that is, all activities of a student can be recorded with a time-stamp as he/she interacts with a computer-based learning environment. Authors in (Cocea and Weibelzahl, 2009) have shown the possibility of detecting and predicting students' disengaged behaviors from their logged data using different data mining techniques. Furthermore, experimental results presented in (Cocea and Weibelzahl, 2009, 2011) and (Tan et al., 2014) have shown generality of a set of attributes that can be collected from most e-learning systems repository, which makes it possible to reuse the developed model or approach with data collected from other learning environments. In particular, exploratory work of Tan et al. (2014) showed that comparing (behavioral) engagement of two groups of students who have worked on different intelligent tutoring systems did not reveal any significant difference. Cocea and Weibelzahl (2011) on the other hand determined the validity of their previously developed engagement detection model using data from a less structured learning management system. These works testify the potential of studying and extracting students' intended learning behaviors from their logged data recorded by computer-based learning environments. A little consensus, however, exists on the representation of the students' engagement behavior.

In this work, we used a previously proposed classification scheme by Maqsood et al. (2019). The scheme transforms students' logged problem-solving actions (during confidence-based assessment) into six engagement/disengagement behavioral patterns, as described in the following.

## 2.1 Mapping Students' Problem-Solving Actions into (Dis)Engagement Behavioral Patterns

Confidence-based assessment is a two-dimensional assessment paradigm that takes students' confidence level with each submitted answer. This additional "confidence" measure in combination with student response's correctness (which could be either correct or incorrect) derives four confidence-outcome categories; by following (Hunt, 2003) and (Vasilyeva et al., 2008) we have: high confidence-correct response (HCCR), low confidence-correct response (LCCR), high confidence-wrong response (HCWR), and, low confidence-wrong response (LCWR). These distinct categories capture a discrepancy between students' expected and actual performance – a *gap* that can be addressed using correct information offered to students through feedback in a computer-based assessment system (Maqsood and Ceravolo, 2019).

The pre-mentioned distinct confidence-outcome categories are defined in terms of varied knowledge regions introduced by Hunt (2003). That is, HCCR shows mastery of a student in the subject domain; LCCR depicts doubt or hesitation about one's knowledge; HCWR means that the student has misconceptions, and LCWR shows unknowing knowledge state of a student. In this respect, authors in (Maqsood et al., 2019) identified that seeking or no-seeking a corrective (or task-level) feedback followed by an answer belonging to a specific confidence-outcome category, can lead to different engaged and disengaged behaviors of the students during assessment.

As intuition suggests, previous results of (Maqsood and Ceravolo, 2019) showed that students' feedback-seeking behavior is positively correlated with wrong answers given with either confidence level. Therefore, the classification scheme proposed by Maqsood et al. (2019) only considers feedback seeking behavior for incorrect responses to differentiate between students' engagement or disengagement during assessment. Table 1 show the complete mapping of students' logged problem-solving actions into corresponding engagement/disengagement behaviors. In the following, we precisely explain the theoretical reasoning underlying the six distinct (dis)engagement behavioral patterns (see last column of the table).

In Table 1, the first two rows contain answers belonging to HCCR and LCCR which respectively represent students' correct responses given with high and low confidence. As mentioned earlier, feedback-seeking action has no correlation with correct responses given with either confidence level (Maqsood and Ceravolo, 2019); therefore, only a single engagement behavioral pattern is defined for each category of response, namely: "High Knowledge" (HK) and "Less Knowledge" (LK).

On the other hand, students' different reactions to corrective feedback (i.e., seeking or no-seeking) in case of wrong responses given with high and low confi-dence derive four distinct engagement and disengagement behavioral patterns. That is, seeking corrective feedback after a high confidence wrong response (HCWR) is interpreted as an engaged behavior of a student who is trying to fill the knowledge gap that occurred as a misconception or discrepancy between his/her expected and actual knowledge – thus, the engagement behavioral pattern is named as "Fill-knowledge Gap" (FG). And, if a student does not perform feedback-seeking action after a HCWR, it is considered as that the student did not attempt to repair the knowledge gap(s); the corresponding disengagement behavioral pattern is called "Knowledge Gap" (KG).

**Table 1:** Mapping of the students' logged problem-solving actions into six (dis)engagement behavioral patterns, as defined in Maqsood et al. (2019). In the Confidence-Outcome Category (first column) – the first two letters, HC or LC, respectively specify a student's high or low confidence level associated with a submitted answer. The last two letters, CR or WR, in a Confidence-Outcome Category specify the correct or wrong response submitted by a student, respectively.

| Confidence-Outcome Category | Student Response to Corrective Feedback | New label for (Dis)Engagement Behavioral Pattern |
| --- | --- | --- |
| HCCR | Feedback Seek (FS) or | High Knowledge (HK) |
| LCCR | Feedback No-Seek[a] | Less Knowledge (LK) |
| HCWR | Feedback Seek (FS) | Fill-knowledge Gap (FG) |
|  | Feedback No-Seek | Knowledge Gap (KG) |
| LCWR | Feedback Seek (FS) | Learn (LE) |
|  | Feedback No-Seek | Not Interested (NI) |

[a]No label is stored for this activity in the traced logs, so it is considered by the absence of FS activity after each submitted problem.

A low confidence wrong response (LCWR) reflects the *unknowing* knowledge state of a student, and therefore, seeking feedback in this case means that a student is trying to learn something; hence this engagement behavioral pattern is referred to as "Learn" (LE) in (Maqsood et al., 2019). Since, the corrective (task-level) feedback was only available to students for the answered questions, a student who does not know the answer to a question may submit a low confident wrong response to see the correct answer and/or detailed explanation. Yet, this behavior reflects that the student attempts to learning something; as rightly captured by the LE behavioral pattern. However, a student who does not perform feedback-seeking activity followed by a LCWR, is considered as showing disengagement during assessment – therefore, the corresponding disengagement behavioral pattern is called as "Not Interested" (NI).

We believe that this classification scheme represents students' active or inactive involvement in confidence-based assessment at varied levels by mapping their problem-solving actions to six (dis)engagement behavioral patterns. Furthermore, the data analysis performed in (Maqsood et al., 2019) showed that the proposed scheme is quite reasonable to distinguish behaviors of high and low performance students as determined from their actual performance in a class. In this research work, we utilized this classification scheme and transforms students' logged problem-solving activities at low-level into corresponding behavioral patterns, details are provided in Section 4.1.

## 2.2 Model-based Clustering

Markov chain is primarily an efficient method to model sequential data and make predictions. However, student engagement is not a stable factor and is subject to change over time (Joseph, 2005), therefore, striving for a single *best* model to represent students' behaviors is not adequate. Hence, to capture dynamic behaviors reflecting student (dis)engagement during assessment, we decided to perform model-based clustering which is a probabilistic method and results in a set of $K$ mixture models (or clusters). All observations belong to multiple clusters with different probabilities and each mixture component represents a different data distribution through a Markov chain. Hansen et al. (2017) also insisted on the use of mixture Markov chains to model sequential traces of students which have the capability to capture *drift* in students' behaviors through different mixture components. Keeping in view the finding of (Cohen and Beal, 2009) which shows that the next action pattern of a student depends more likely on the previous pattern and not much on earlier patterns, we select first-order Markov chains to model and predict students' likely engagement behaviors. Thus, each mixture component is represented by a first-order Markov chain.

Statisticians refer to model-based clustering as a mixture model of $K$ components (Cadez et al., 2003) and, in the literature, the terms are often used interchangeably. However, model-based clustering requires an additional step than just finding a finite mixture model, that is, to assign each sequence to its appropriate cluster from $K$ mixtures based on a pre-specified rule (Melnykov et al., 2010). The most commonly used approach is a *Bayes' decision rule* which assigns a sequence to the mixture with maximum (log-)likelihood.

Expectation-Maximization (EM) algorithm is a well-known iterative procedure to determine a finite mixture model by maximizing the likelihood of observing a complete dataset. More precisely, the mixture modeling framework assumes that each sequence *s* is generated by one of the $K$ component distributions, however, its true membership label is unknown (Melnykov, 2016). EM algorithm aims to incorporate these missing labels. That is, given some observed data *y*, EM tries to find a model $\vartheta \in \Theta$ with maximum (log) likelihood estimation (MLE) (Gupta et al., 2011), where $\Theta$ is the symbol of parameter values. Formally:

$$\hat{\boldsymbol{\theta}}_{MLE} = arg\ max_{\theta \in \Theta}\ \log p(y|\vartheta) \tag{1}$$

In order to find such a model, the EM algorithm iterates over the following two steps until it reaches convergence (or some stopping criterion).

- Expectation (or E) step: estimates the conditional expectation of complete-data log-likelihood function given the observed data.
- Maximization (or M) step: finds the parameter estimates to maximize the complete-data log-likelihood from the E-step.

Finding an optimal *global* maxima is challenging for EM and it usually ends up with one of the best *local* maxima. However, the initialization of the algorithm parameters plays a critical role in finding an optimal solution (Michael and Melnykov, 2016; Hu, 2015). To cluster multivariate categorical data, EM algorithm requires the following three parameters to get started:

1. Number of mixtures ($K$).
2. Initial transition matrices for $K$ mixtures.
3. Initial weights of $K$ mixtures.

Like K-means, the EM algorithm also requires a prior number of mixtures to be defined by the user which is one of the challenging problems for researchers. However, model-based clustering has the advantage of being supported by formal statistical methods to determine the number of clusters and model parameters (Magidson and Vermunt, 2002). The two most commonly used methods which are based on 'information criterion' to select the optimal value of $K$ are Bayesian Information Criterion (**BIC**) (Schwarz et al., 1978) and Akaike Information Criterion (**AIC**) (Akaike, 1998). Both methods penalize complex models, thus, the models with the lowest **BIC** and **AIC** scores are better. The primary difference between both measures is that **BIC** penalizes heavily in contrast to **AIC**.

*EM algorithm variants:* There are many variants of the EM algorithm available in the current literature. We selected the two most basics one for our experimental work, that is, the original EM and emEM.

- EM (Dempster et al., 1977) — the original EM algorithm in which initialization is performed randomly. The standard EM algorithm initializes the initial transition matrices for $K$ mixtures randomly where $K$ is given by the user.
- emEM (Biernacki et al., 2003) — a variant of the EM algorithm in which the EM algorithm is also run in the initialization phase for a given $K$, as reflected by the prefix *em*. The best model is then picked as the starting point (or initial model) followed by the actual EM algorithm.

All the mixture components are assigned an equal initial weight (i.e. $1/K$) in both EM and emEM algorithm.

In this exploratory work, we also proposed a new initialization method called K-EM which uses the results of K-means clustering performed on students' problem-solving actions. The details of our approach is given in the next section.

## 3 K-EM: Proposed Initialization method for the EM algorithm

As mentioned before, initializing the EM algorithm using partitioning obtained through K-means or Hierarchical clustering method is referred to as a practical solution (Gupta et al., 2011; Michael and Melnykov, 2016) to avoid *local* maxima. According to Gupta et al. (2011), performing a preliminary cheaper clustering like K-means or Hierarchical for initializing the EM algorithm is expected to give better results than random assignment. In their work, this approach is used for Gaussian Mixture Model (GMM) where clusters' means and covariance matrices are taken from the K-means results. Hu (2015) used hierarchical clustering for the initialization of the EM algorithm for finding model parameters for GMM. However, this is not straightforward in case of multivariate categorical data. In the following, to describe our proposed K-EM method, first we discuss in detail that how K-means is performed on our datasets and then how the retrieved results are used to initialize the EM algorithm.

In our problem, we have two datasets containing students' logged interactions from computer-based assessment systems. The logged students' problem-solving actions are mapped to six different (dis)engagement behavioral patterns as mentioned in Section 2.1. Our intention to use K-means on this kind of data came from a previous work (Maqsood et al., 2019) which resulted into distinctive clusters of similar traces representing students' (dis)engagement behaviors. An argument may arise here that another variant of the K-means clustering called K-modes (Huang, 1998) is more suitable for categorical data, which defines the similarity between two sequences based on matching elements. But, our datasets contain traces of different lengths and we used each behavioral pattern's proportional count to represent a trace as done in (Maqsood et al., 2019). This data transformation is represented by Eq. (2), which computes the proportional count for each behavioral pattern $p_i \in P$ per trace, where $P = \{HK, LK, FG, LE, KG, NI\}$.

$$\frac{\sum_{i \in P} p_i}{\text{Trace length}} \quad (2)$$

In the following, we present an example to illustrate this data transformation for a trace.

> **Example:** Let's take a sample trace of length four: <HK, FG, KG, HK>– each element represents a mapping of a student's problem-solving actions into corresponding (dis)engagement behavioral pattern as described in Section 2.1. The proportional count for each behavioral pattern in this sample trace using Eq. (2) is: HK=0.5; LK=0; FG=0.25; LE=0; KG=0.25; NI=0. Therefore, the initial sample trace after this data transformation would become <0.5, 0, 0.25, 0, 0.25, 0>(behavioral patterns as given in the set *P* are replaced by their proportional count in a specific trace). All the traces in both datasets were converted into patterns' proportional count in a similar fashion.

**Table 2:** The proposed K-EM method

---

1. Perform K-means clustering on input data (as described in the text).
2. Use the results from Step 1 to initialize the EM algorithm in the following manner.
   (a) Set the number of mixtures ($K$) equal to the number of clusters ($K'$) used in K-means algorithm.
   (b) Construct a first-order Markov chain for each resultant cluster ($C_{k'}$) containing $T_{k'}$ traces. Use these transition matrices as initial transition matrix for respective $K$ mixture components.
   (c) Weights of $K$ mixtures are set to the ratio of the number of traces in each respective obtained cluster, that is, $w(C_k) = T_{k'} / \sum_{i=1}^{k} T_i$.
3. Run the usual EM algorithm.

---

Consequently, traces containing similar distribution of distinct behavioral patterns were grouped together in a same cluster using Euclidean distance. The results of K-means clustering are then used to initialize the EM algorithm – leading to our proposed method, K-EM, as given in Table 2. Students depict diverse problem-solving actions during assessment which makes the problem even more challenging to find a suitable representation of them. Also, the datasets collected in educational studies are usually in small to medium sizes, we expect that our proposed data-specific initialization method will result into better mixture Markov models capturing the students' varied (dis)engagement behaviors.

## 4 Methodology

In this section, first we described the two real datasets used in our experimental work along with the design of the two studies conducted to collect these datasets. Then, we present our methodology for constructing and evaluating the mixture Markov models.

### 4.1 Datasets Description

In this study, we have used two real datasets namely, Dataset1 and Dataset2, containing students' logged interactions with computer-based assessment systems. The first experimental study involved 94 freshmen from the National University of Computer and Emerging Sciences, Pakistan, while the second experiment was conducted with 210 undergraduate students of the Universitá degli Studi di Milano, Italy.

*(a) First experimental study design:* In the first study, three sessions of 40-45 minutes each were conducted in different weeks and students were given six (code tracing) problems per session on a computer-based assessment system.

A task-level detailed feedback (referred to as *corrective feedback* in Section 2.1 for each question was provided to a student upon request by the computer-based assessment system. The corrective feedback shows correct answer along with its explanation; see Maqsood and Ceravolo (2019) for details of the tool and study material used in the first experimental study. The experiment was conducted in a self-assessment setting, that is, no time limit was specified for any question and there was no impact on a student's course records based on his/her participation and/or performance in this study. Students were asked to solve as many questions as they can in the given time and specify their confidence level (as high or low)[1] before submitting a solution. In fact, two submit buttons ('High confidence submit' and 'Low confidence submit') were available (on student portal) so that students can make a conscious choice of their confidence level for each answer; see Maqsood and Ceravolo (2019) for details of the tool and study material used in the first experimental study. The students' logged interactions collected from this first study are referred to as Dataset1.

*(b) Second experimental study design:* In the second study, multiple choices questions were given using a computer-based assessment system to evaluate students' comprehension of given flow diagrams. The computer-based assessment tool used in this study also offered corrective feedback for each question upon a student's request. This study was also conducted for students' self-assessment purposes, however, with relatively different settings. The class teacher uploaded 39 multiple choice questions related to basic concepts of an introductory programming course. More specifically, 13 different exercises were uploaded with code flow diagrams. Each exercise contained 3 multiple choice questions, each on a separate page. Students were asked to use the tool for their self-assessment and preparation of the final examination. As before, students were required to specify a confidence level (as high or low) with each submitted response using a dedicated submit button (i.e. *High confidence submit* and *Low confidence submit* ). The students' logged interactions collected from this second study are referred to as Dataset2.

The purpose of conducting the both experimental studies was to collect data for determining and analyzing students' (dis)engagement behaviors during confidence-based assessment. Computer-based assessment tools used in both studies recorded students' problem-solving actions along with their timestamp. The assessment model used for designing the tools and data collection is given in Maqsood and Ceravolo (2018).

Since, students were free to solve any number of problems in both experimental studies, their recorded problem-solving actions per Login-Logout session were of different lengths. The collected datasets contain sequentially ordered activities for each Login-Logout session of respective students. We refer to a Login-Logout session containing a student's problem-solving actions as a "trace" in the following text. During data pre-processing, we removed traces of length 1 as we needed to compute transition matrices of traces and make predictions, which is impossible for a single event trace. Thus, we are left with the traces of 92 students in Dataset1 and 185 students in Dataset2. Table 3 contains a summary of the remaining datasets.

---

[1] We used binary scale for confidence measurement instead of a more complex rating (e.g. percentage rating between 0-100, Likert scale response, etc.), which may confuse students in estimating their confidence about solution's correctness (Petr, 2000; Vasilyeva et al., 2008).

**Table 3:** Summary of solved problems in both datasets

| Data | Number of solved problems | | | |
| --- | --- | --- | --- | --- |
| | Minimum | Maximum | Average | Total |
| Dataset1 (92 st.; 197 traces) | 2 | 6 | 5 | 1033 |
| Dataset2 (185 st.; 348 traces) | 2 | 39 | 17 | 5771 |

**Table 4:** Students' sample traces with trace lengths between minimum 2 and maximum 6; all the behavioral patterns belong to the set $P = \{$HK, LK, FG, LE, KG, NI $\}$, and, separated by a hyphen "-"

| | |
| --- | --- |
| Trace1: | HK-HK-LK-HK-LK |
| Trace2: | HK-HK-FG |
| Trace3: | KG-KG-LE-KG |
| Trace4: | HK-HK-HK-FG-FG-FG |
| Trace5: | HK-FG-KG-FG |
| Trace6: | LE-LE-LK-LK-LK |
| Trace7: | NI-NI |
| Trace8: | HK-HK |
| Trace9: | LK-HK-FG-FG-LK-LK |
| Trace10: | FG-HK-HK-LK |

**Table 5:** Distribution of behavioral patterns in the two datasets, N is the total count of behavioral patterns in a dataset (value in the parentheses show percentage of a corresponding behavioral pattern)

| Behavioral Pattern | HK | LK | FG | LE | KG | NI |
| --- | --- | --- | --- | --- | --- | --- |
| Dataset1 (N=1033) | 421 (40.8%) | 35 (3.4%) | 363 (35.1%) | 81 (7.8%) | 117 (11.3%) | 16 (1.6%) |
| Dataset2 (N=5771) | 2047 (35.5%) | 1769 (30.7%) | 671 (11.6%) | 1094 (19%) | 52 (0.9%) | 138 (2.4%) |

All the problem-solving actions contained in both datasets are transformed into respective discrete engagement and disengagement behavioral pattern (as mentioned in Section 2.1). Table 4 shows 10 sample traces having trace lengths between 2 and 6 (representing respectively the minimum and maximum number of problem-solving actions encoded into their corresponding (dis)engagement behavioral pattern from the set $P$). Table 5 shows the distribution of each behavioral pattern in both datasets.

Clearly, both datasets have class imbalance problem which can make it difficult to predict the infrequent behavioral patterns due to insufficient data for model training. More specifically, Dataset1 is dominated by behavioral patterns that belong to high confidence, that is, HK and FG – which respectively represents high knowledge and positive engagement of the students (as described in Section 2.1. And, behavioral patterns belonging to low confident students, that is, LE, LK and NI are below than 9% of the total behavioral patterns in the Dataset1. Whereas, Dataset2 is dominated by correct knowledge attained with high and low confidence, as represented by HK and LK behavioral patterns, respectively. And, this dataset majorly lacks reasonable representation of behavioral patterns representing disengagement of the students, i.e., KG and NI (both are below than 3% of the total behavioral patterns in the Dataset2).

## 4.2 Methodology for Constructing and Evaluating the Mixture Markov Models

As mentioned in Section 1, the two objectives of this research work are: (1) to construct mixture Markov models for finding a suitable representation for the students' traces containing varied (dis)engagement behavioral patterns; and, (2) to predict students' future behavioral pattern given their previous history. Our methodology for constructing and evaluating the mixture Markov models is shown in Fig. 1. At the top, the input data is shown that contains students' sequential traces of varying lengths, wherein each trace comprises of (dis)engaged behavioral patterns. The upper part of Fig. 1 is labeled as "Data splitting". Below that, the left-half and right-half sides are respectively labeled as "Phase-I: Model Construction" and "Phase-II: Model Evaluation" for separating the two phases clearly. In the following, we present details of these three sub-phases of our methodology.

*4.2.1 Data splitting*

For constructing and evaluating our mixture Markov models, we split the datasets using student-level 5-folds cross-validation. The notion of *student-level* is a better alternative of *student-stratification* in Educational Data Mining because it separates the students between *train* and *test data* (Pelánek, 2018). Also, the student-level data splitting relates to the real-world scenario where we want to train a model on old students' data and then use that model for predicting behaviors of future students.

With 5-folds student-level data splitting, students in both datasets were randomly assigned to 5-folds. Once the student-level folds have been created for both datasets, we retained them to run different algorithms for making justifiable comparisons and analyses.

*4.2.2 Phase-I: Model Construction*

The data from i-1 folds is then treated as *train data* and is used in the Phase-I for model construction. Using the train data as input, we obtained $K$ mixture (or clusters) by running the EM, emEM and K-EM methods. We also constructed a non-mixture baseline model for comparisons, which is described in the Section 5.

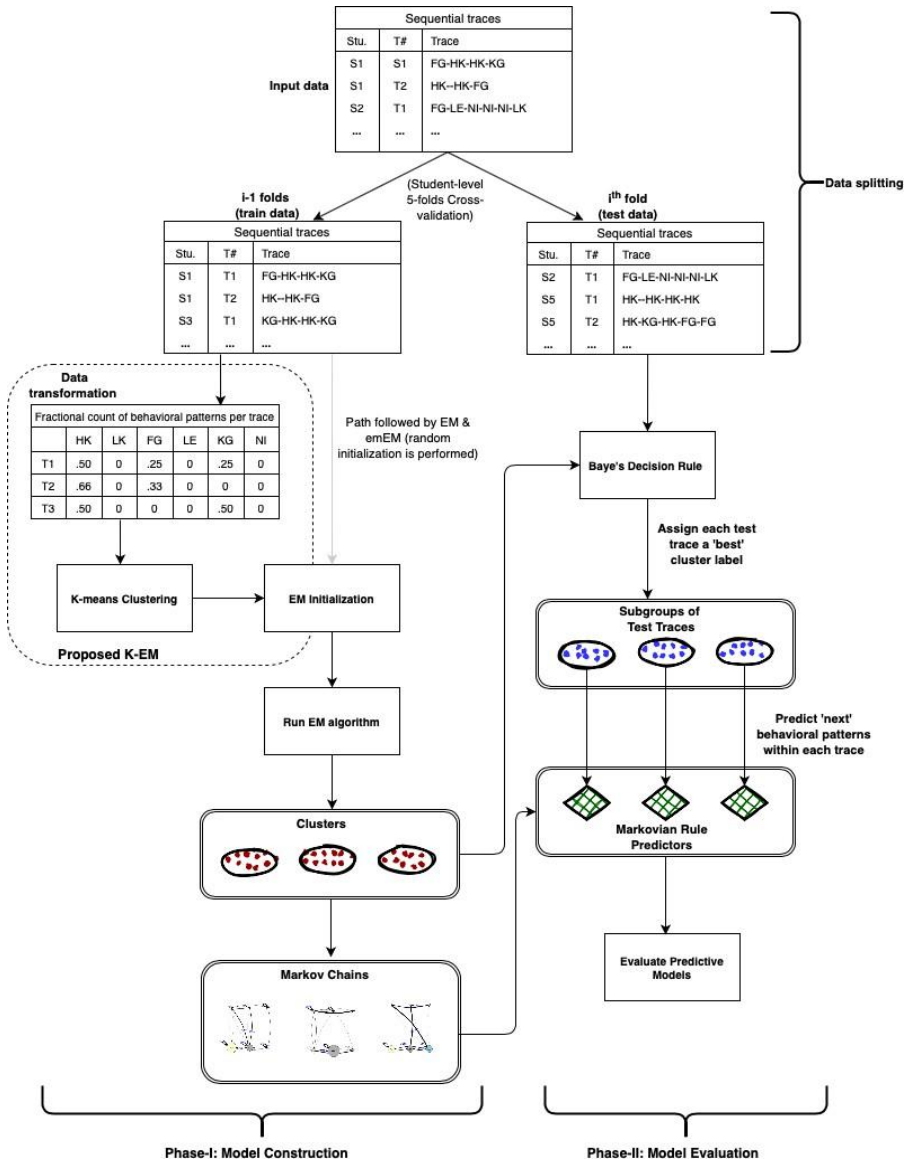

**Fig. 1:** Methodology for constructing (Phase-I) and evaluating (Phase-II) the mixture Markov models using student-level 5-folds cross validation

The resulted *K* mixtures are shown as *Clusters* in Fig. 1, each comprises of similar traces of the students in the *train data*. For each mixture, we con-structed a corresponding first-order Markov model, shown as *Markov Chains* in the figure.

### 4.2.3 Phase-II: Model Evaluation

The next important step is to evaluate the constructed mixture Markov models from Phase-I. For this purpose, the data from i[th] fold is treated as *test data*. The *test data* contains traces of the students different from those in the *train data*. Since, we got $K$ models from the Phase-I, our first task in the Phase-II is to find the most suitable mixture label for all the *test data* traces so that we can use the corresponding trained model for making predictions in a trace. This is the very usual approach in clustering based classification designs wherein a clustering method is used first to create $K$ number of models and then the most suitable model amongst those is used for test data classification, for example, see Lopez et al. (2012). Also, note that the resultant first-order Markov chains which will be used for making predictions of the students' future behavioral patterns, are constructed already in the Phase-I and they have not seen the *test data* yet. In the followings, we describe our approach for Bayes' decision rule for labeling test traces, model for predicting students' future behavioral patterns, and, evaluation metrics that we have used for performance estimation of the mixture Markov models, sub-modules shown in Fig. 1 – Phase-II.

*(a) Bayes' decision rule for labeling test traces:* Given the $K$ mixtures generated earlier in Phase-I, we estimated the posterior probability of each test data trace $t$. Then, we used the *Bayes' decision rule* [2] to find the most suitable mixture label for a trace based on the maximum posterior probability. After performing this step, all the traces in the *test data* got a mixture label; thus, we can imagine that there will be some similar test traces which are estimated to belong to a specific mixture. We have shown this step in Fig. 1 as "*labeling test traces*" which produces subgroups of similar test data traces.

*(b) Predicting students' future behavioral patterns:* Next, we wanted to predict students' future behavioral patterns using our pre-constructed mixture Markov models from Phase-I (shown at the bottom of Fig. 1 as first-order Markov chains). Also, prediction is a mechanism for validating the developed learner models (Desmarais and Baker, 2012), which in our context represent students' engagement/disengagement behaviors by first-order Markov chains. Markov chains serve dual purposes of modeling and predicting sequentially ordered activities. With first-order Markov chain, we make the *Markovian* assumption that a student's future behavioral pattern $b_{i+1}$ is dependant on his/her current behavioral pattern $b_i$ only and not on the previous history. That is:

$$P(b_{i+1}|b_1, b_2, ..., b_i) = P(b_{i+1}|b_i) \qquad (3)$$

---

[2] We have provided the R code implementation of this step at GitHub; https://github.com/r-maqsood/Mixture-Markov-Models-R.

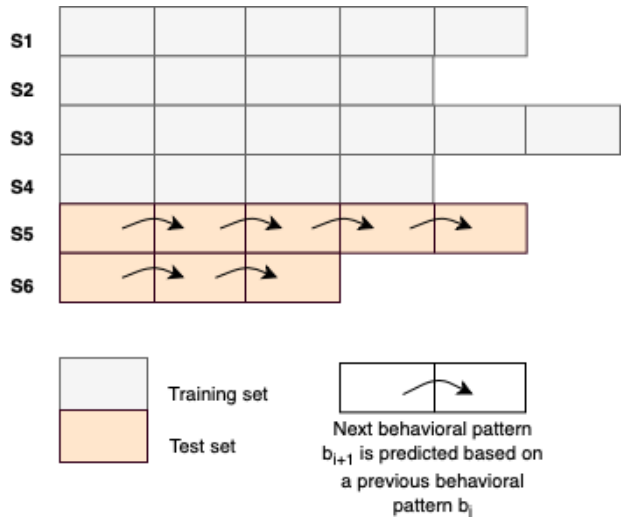

**Fig. 2:** Dynamic behavioral pattern predictions made for students' traces. Each future behavioral pattern $b_{i+1}$ is predicted based on the current behavioral pattern $b_i$ using *Markovian* assumption, and hence, no prediction is made for the first behavioral pattern in a trace. This figure is a reproduced version of Fig. 5(d) from Pelánek (2018) with minor changes.

Here, $P$ represents the conditional probability of an event $b_{i+1}$ given some previous event(s).

Fig. 2 illustrates our approach for making predictions at student trace level which is referred to as "dynamic prediction" by Pelánek (2018). As shown in the figure, given a model trained on students' sequential traces in the *train data*, we can use it to make predictions for student in the *test data*. And, after observing each new event in a student trace, the prediction is updated. In our case, we used first-order Markov chain to represent sequential traces in the *train data* and predict a student's next behavioral pattern using Eq. 3. This means that for a student trace of length $l$ in the *test data*, we make $l-1$ predictions. Note that no prediction is made for the first behavioral pattern since there is no preceding behavioral pattern.

We remind here that we have already constructed $K$ number of first-order Markov chains in the Phase-I (see Fig. 1), each corresponding to a resultant mixture from a specific variant of the EM algorithm used in this research study. Cadez et al. (2003) also showed that a mixture of first-order Markov chains is different than a simple (or non-mixture) first-order Markov chain and that making predictions with the prior approach resulted into a better accuracy.

*(c) Performance evaluation metrics:* In order to estimate the performance of different mixture Markov models, evaluation metrics that we have used are listed in Table 6. For our multi-class predictive models obtained for different clusters, we used prediction accuracy, precision, recall and F1 score for performance evaluation. We also computed the number of iterations taken by an algorithm for model training (Phase-I, Fig. 1) since a faster variant of the EM algorithm can be a matter of choice in the case of large datasets.

Table 6: Performance metrics used for model evaluation

| Metric | Description |
|---|---|
| Macro Acc.$_t$ | Macro accuracy: prediction accuracy computed at students' trace-level; (Eq. 4) |
| Micro Acc. | Micro accuracy: prediction accuracy computed using complete test data; (Eq. 5) |
| Precision$_{wt.}$ | Weighted (macro) precision; (Eq.6) |
| Recall$_{wt.}$ | Weighted (macro) recall; (Eq. 7) |
| F1$_{wt.}$ | Weighted (macro) F1 score; (Eq. 8) |
| Iterations | Number of iterations taken by an algorithm for model training using train data |

The usual *micro* performance metrics (i.e., micro-accuracy, micro-precision, micro-recall, micro-F1) are computed for complete test data which treats all the classes equally (e.g., see Eq. 5). In case of class imbalance problem, a more dominant class(es) can overshadow the rare or less frequent class(es), leading to incorrect performance measures. Since, both of our datasets have class imbalance problem, we focused on macro and weighted (macro) versions of these metrics as mentioned in Table 6. However, we report both micro and macro prediction accuracy measures for completeness. These metrics are computed for predictive models corresponding to $K$ mixture Markov models. To summarize the performance of an algorithm ran in a 5-folds cross-validation setting, we computed weighted average of $K$ clusters (we simply refer to it as mean (M)) and standard deviation (SD) of all these metric (as reported in the next section).

$$\text{Macro Acc.}_t = \frac{\text{No. of correct predictions per trace}}{\text{Trace length}} \quad (4)$$

$$\text{Micro Acc.} = \frac{\text{No. of correct predictions}}{\text{Total predictions}} \quad (5)$$

$$\text{Precision}_{wt.} = \sum_{i \in C} \text{Precision}(\text{Class}_i) \times \text{Weight}(\text{Class}_i) \quad (6)$$

$$\text{Recall}_{wt.} = \sum_{i \in C} \text{Recall}(\text{Class}_i) \times \text{Weight}(\text{Class}_i) \quad (7)$$

$$\text{F1}_{wt.} = \sum_{i \in C} \text{F1}(\text{Class}_i) \times \text{Weight}(\text{Class}_i) \quad (8)$$

In the above equations, *C* is the total number of classes which represent the six behavioral patterns in our datasets. Weight(Class$_i$) is the ratio of the number of behavioral patterns that belong to Class *i*.

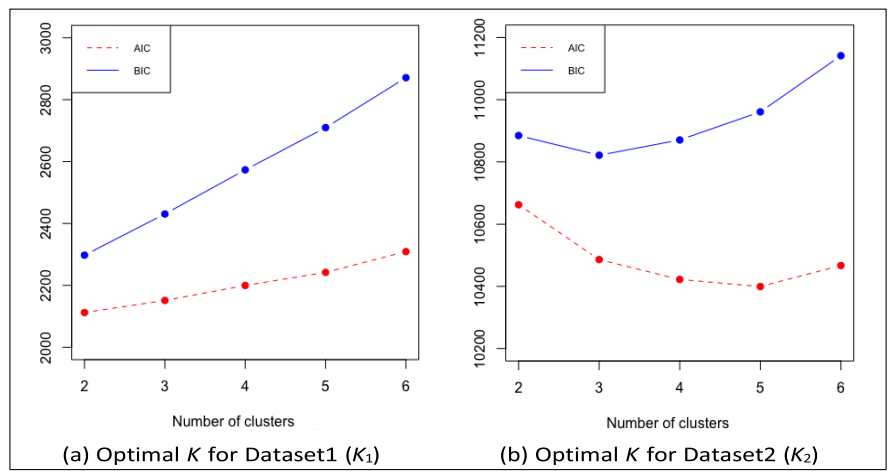

(a) Optimal $K$ for Dataset1 ($K_1$)  (b) Optimal $K$ for Dataset2 ($K_2$)

**Fig. 3:** Models comparison using AIC and BIC scores to determine the optimal number of mixtures $K$ for EM and emEM algorithms – (a) Dataset1: $K_1$=2, (b) Dataset2: $K_2$=3

## 5 Experimental Setup

All the experiments related to model-based clustering in this work were performed using ClickCluct package of R (Melnykov, 2016), which provides an implementation of the emEM algorithm. The algorithm converges (or stops) if the difference between the log-likelihood of two subsequent iterations is less than $1e - 10$. We used the same stopping criterion for the EM and K-EM algorithms, and, modified the existing code to implement these two variants. The following two sub-sections explain the parameters used to construct mixture Markov models using the three variants of the EM algorithm. More specifically, we provide details of how the three algorithms, that is, EM, emEM and K-EM (our proposed method) initializes the three initial parameters of the Expectation-Maximization algorithm (as mentioned in Section 2.2. Moreover, we also constructed baseline models for both datasets as described in Section 5.3.

5.1 Initial parameters for the EM and emEM algorithms

*(a) Number of mixtures (K):* For determining the optimal value of $K$ for both EM and emEM algorithms, we computed BIC and AIC scores for both datasets using models of different number of mixtures, see Fig. 3 ($K$ is on the horizontal axis).

For Dataset1, Fig. 3(a) shows that the BIC and AIC scores increase with an increasing $K$ value and both measures suggest that 2 is the optimal number of clusters. Whereas, for Dataset2, BIC and AIC measures disagree on the optimal value of $K$ (see Fig. 3(b)); that is, the lowest BIC score is achieved at $K = 3$ and the lowest AIC score is at $K = 5$. In such a situation, the BIC-preferred value can be taken as a minimum value of $K$ and the AIC-preferred value as a maximum $K$ value; and, any model can be picked within this range (preferably based on some other criteria) (Dziak et al., 2019). For Dataset2, the range for the optimal number of mixtures $K$ is 3 and 5, and we picked $K = 3$ arbitrarily. Thus, the EM and emEM algorithms were applied on Dataset1 and Dataset2 using $K_1=2$ and $K_1=3$, respectively.

*(b) Initial transition matrices for K mixtures:* For initial transition matrices, the EM algorithm uses random values. While, the emEM algorithm runs the EM algorithm in the initialization phase and finds the approximate values for transition matrices as the starting point.

*(c) Initial weights of K mixtures:* All the mixture components are assigned an equal initial weight (i.e. $1/K$) in both EM and emEM algorithms.

5.2 Initial parameters for the K-EM algorithm

*(a) Number of mixtures (K):* To select the optimal number of clusters $(K')$[3] for both datasets, we used the NbClust method of R which uses 30 different well-known indices for approximation, including: Cindex, CH index, Beale index, DB index, Silhouette index, Dunn index, etc. (see Charrad et al. (2012) for details). The NbClust methods retrieves the best value of $K$ using maximal voting between all the indices. For our datasets, we got 4 and 2 as the optimal values of $K'$ for Dataset1 and Dataset2, respectively.

The *elbow graph*, which is a very popular approach to visualize the optimal value of $K$ for K-means algorithm, is shown in Fig. 4 for both datasets. We can see that the optimal $K'$ determined for both datasets using NbClust method (i.e. $K'_1=4$ for Dataset1 and $K'_2=2$ for Dataset2) are indeed good choices as indicated by minimum within-cluster sum of squares.

Next, using the values of $K'_1$ and and $K'_2$, we run the K-means algorithm on both datasets as described in Section 3. The K-means algorithm was executed for 15 iterations with 25 initial points – which is often a recommended approach for finding better clusters by repeating the algorithm with different initial centroids. The results of K-means were then used to initialize the EM algorithm as mentioned earlier in Section 3.

*(b) Initial transition matrices for K mixtures:* Using the results of K-means clustering algorithm, initial transition matrices for the EM algorithm were initilized as mentioned in Table 2, 2(b).

*(c) Initial weights of K mixtures:* Initial weights of the $K$ mixtures were set to the number of traces belonging to corresponding clusters obtained by the K-means algorithm; see Table 2, 2(c)

---

[3] We referred to the value of $K$ used in K-means algorithm as $K'$ to differentiate it from the optimal value of $K$ used for the EM and emEM algorithms.

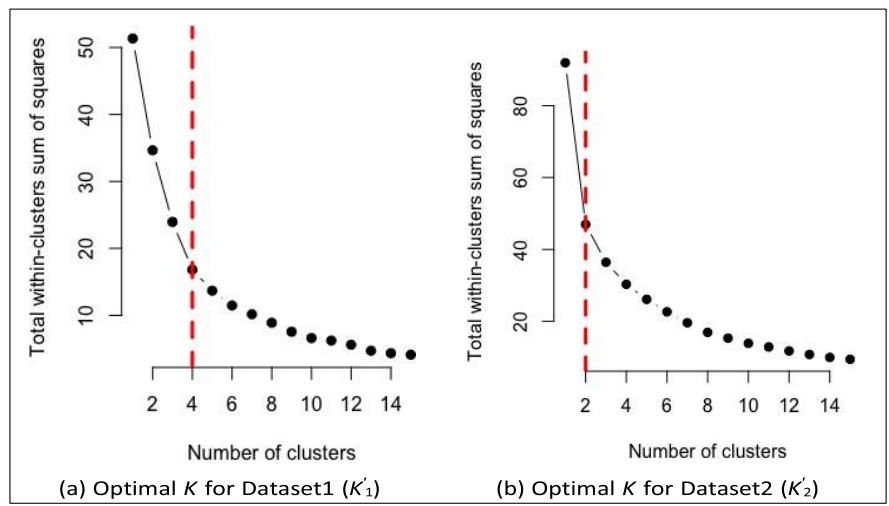

(a) Optimal *K* for Dataset1 ($K'_1$)  (b) Optimal *K* for Dataset2 ($K'_2$)

**Fig. 4:** Elbow graphs showing the optimal number of clusters to be used in the K-EM method – (a) Dataset1: $K'_1$=4, (b) Dataset2: $K'_2$=2

## 5.3 Baseline Model

Our baseline models for both datasets are non-mixture models, that is, no mixtures or clusters are created and hence only a single model is constructed for each dataset in each iteration of 5-folds student-level cross-validation. Given the heterogeneous nature of our datasets, it is expected that the three model-based clustering methods will have better prediction accuracy than the corresponding non-mixture baseline models of both datasets.

## 6 Results Analyses and Discussion

In this section, we present the results of our experimental work performed using the three variants of the EM algorithm – EM, emEM, and K-EM, and baseline models for both datasets; details of their initial parameters settings are given in the previous section. The evaluations metrics that we have used for models' performance estimation are mentioned in Table 6. We provided (weighted) mean and standard deviation of each evaluation metric computed over 5-folds student-level cross-validation.

### 6.1 Results Analyses for Dataset1

Table 7 show the results of Dataset1, the three variants of the EM algorithm were ran using the corresponding optimal *K*, see the values of $K_1$ and $K'_1$, respectively in Fig. 3 and Fig. 4.

**Table 7:** Comparison of student-level 5-folds cross validation results for Dataset1 *(92 students; 197 traces; models constructed as described in Section 5; EM and emEM are run with $K_1$=2; K-EM with $K'_1$=4 mixtures; Baseline Model does not contain any cluster)*

|  | Baseline Model | | EM | | emEM | | K-EM | |
| --- | --- | --- | --- | --- | --- | --- | --- | --- |
|  | Mean | SD | Mean | SD | Mean | SD | Mean | SD |
| Macro Acc.$_t$(%) | 53.48 | 2.4 | 56.61 | 2.7 | 55.66 | 4.7 | 62.26 | 3.7 |
| Micro Acc.(%) | 53.09 | 3.9 | 54.90 | 2.8 | 54.03 | 6.4 | 60.98 | 4 |
| Precision$_{wt.}$(%) | 51.72 | 5.3 | 48.19 | 6.5 | 42.69 | 9 | 55.35 | 4.7 |
| Recall$_{wt.}$(%) | 53.09 | 3.9 | 54.90 | 2.8 | 54.03 | 6.4 | 60.98 | 4 |
| F1$_{wt.}$(%) | 52.23 | 4.6 | 49.28 | 3.2 | 46.40 | 7.4 | 53.93 | 3.3 |
| Iterations | – | – | 86.2 | 44.2 | 55 | 54.3 | 147.8 | 58 |

SD = Standard Deviation

In Table 7, we can see that the prediction accuracy (both macro and micro) obtained by all variants of the model-based clustering achieved better results than the non-mixture Baseline model, which contained a single predictive model for the whole *train data*. Though, micro accuracy results for the EM and emEM method are quite similar to that of the Baseline model. Since, the micro accuracy is computed at a global level, that is, it treats all the classes equally; we can expect that some of the very infrequent behavioral patterns (as shown in Table 5) would have impacted this performance measure for the model-based clustering methods which construct $K$ predictive models. So, reporting micro accuracy results for a multi-class imbalanced data is probably not a good choice. For this reason, we will not focus on this metric for future results analyses. Besides this, both randomly initialized methods, i.e., EM and emEM have achieved poor precision and F1 scores in comparison to the Baseline model. Both algorithms have also struggled in achieving good recall in contrast to the Baseline model. Whereas, our proposed K-EM method has overall shown better performance than the Baseline model.

EM and emEM algorithms, both achieved very similar macro accuracy, micro accuracy and recall results; whereas (weighted) precision and F1 scores are better for the EM model. Clearly, our proposed K-EM method has also achieved better results in comparison to the both randomly initialized methods, EM and emEM for Dataset1. However, we remind that our proposed K-EM method is run with $K'_1$=4; whereas the EM and emEM uses $K_1$=2 for Dataset1. This difference in number of mixtures has a clear impact on the number of iterations required for model training. As shown in Table 7, both EM and emEM converges faster than our K-EM method, and it seems reasonable since K-EM has more number of mixtures. However, the variance in the iterations taken by all the algorithms is quite high and we will see shortly if this performance difference is actually meaningful or not.

We could have reported confidence interval (CI) values for the mean results of 5-folds student-level cross-validation shown in Table 7. For example, the CI computed on mean Macro Acc.$_t$ of K-EM, EM and emEM methods. However, those values would not have made much sense for making inference about the mean difference between any two methods (Cumming and Finch, 2005). As mentioned in Section 4, same student-level folds were used for all the algorithms; this allows us to make inference on paired data. Therefore, we computed 95% CI values between the paired difference of each evaluation metric obtained in 5-folds. The results are reported in Table 8 for Dataset1 where we compared our K-EM method with the others; paired difference mean, paired difference standard deviation and 95% CIs are presented. The null value of the CI for the mean difference is zero which means that there is no significant performance difference between the two methods. Thus, the CIs ranges not involving the value of zero are shown in boldface, which indicates a significant performance difference between the two methods with 95% confidence. However, for some metrics/methods we have reported results using 90% CI to show significant difference in the performance, if possible.

Based on the results of Table 8, we can say with 95% confidence that our proposed K-EM method has shown significant performance difference for Dataset1 in comparison to the Baseline and EM methods on the following metrics: Macro Acc.$_t$, Micro Acc. and Recall. However, the two randomly initialized methods – that is, EM and emEM did not show any significant performance difference in comparison to the Baseline model of Dataset1 (even using 80% CI) on any metric[4]. The K-EM has shown significant performance difference with 95% CI in comparison to the emEM method on almost all the metrics (precision results are significant with 90% CI only). The significant difference between Iterations (on last row) show that the emEM converges faster than our proposed K-EM method whereas it had less number of mixtures (i.e., $K_1$=2 for EM and emEM; $K'_1$=4 for K-EM).

To analyze the detailed performance for distinct (dis)engagement behavioral patterns in the resultant $K$ mixture Markov models, we plot summarized confusion matrices of 5-folds student-level cross-validation. Fig. 5 shows confusion matrices heatmap of the resultant $K$ mixtures for Dataset1 using the K-EM algorithm with $K'_1$=4. As we can see, the resultant $K$ mixtures have shown different performance (i.e., true positives, false positive and true negatives) for distinct (dis)engaged behavioral patterns. We remind that the predictions are made using Markovian property (see Eq. 3, which uses the conditional probability of a most recent event to predict a future event – an event in our work represents a specific (dis)engaged behavioral pattern from the set $P = \{$HK, LK, FG, LE, KG, NI$\}$). The three most infrequent behavioral patterns relating to low-confidence in the Dataset1, i.e., LE, LK, and NI (as shown in Table 5) has more false negatives. Whereas, the two most frequent behavioral patterns (i.e, HK and FG) have more false positives which has negatively affected the weighted precision of the resultant Markov models, as shown in Table 7. This problem particularly occurs in the students' traces containing mixed behavioral patterns wherein a student depicted abrupt (dis)engaged behavioral patterns during assessment and hence, our predictive model based on the Markovian property make some incorrect predictions.

---

[4] These results are not reported in Table 8 for conciseness

**Table 8:** Dataset1: 95% confidence interval for paired difference over 5-folds student-level cross-validation. A positive value in the "Diff. Mean" column shows that the K-EM method has better performance in difference means of the two methods; and a negative value meant otherwise (except for the "Iterations" metric where a low value represents a better performance).

|  | Metric | Diff. Mean | Diff. SD | 95% CI |
|---|---|---|---|---|
| K-EM/Baseline | Macro Acc.$_t$(%) | 8.79 | 4.5 | **(3.23, 14.35)** |
|  | Micro Acc.(%) | 7.89 | 4.6 | **(2.21, 13.57)** |
|  | Precision$_{wt.}$(%) | 3.63 | 6 | (-3.81, 11.06) |
|  | Recall$_{wt.}$(%) | 7.89 | 4.6 | **(2.21, 13.57)** |
|  | F1$_{wt.}$(%) | 1.70 | 6.2 | (-5.94, 9.33) |
| K-EM/EM | Macro Acc.$_t$(%) | 5.66 | 2.1 | **(3.06, 8.25)** |
|  | Micro Acc.(%) | 6.08 | 1.8 | **(3.85, 8.30)** |
|  | Precision$_{wt.}$(%) | 7.16 | 11.1 | (-6.58, 20.89) |
|  | Recall$_{wt.}$(%) | 6.08 | 1.8 | **(3.85, 8.3)** |
|  | F1$_{wt.}$(%) | 4.65 | 4.6 | **(0.22, 9.07)\*** |
|  | Iterations | 61.60 | 80.4 | (-38.20, 161.40) |
| K-EM/emEM | Macro Acc.$_t$(%) | 6.60 | 3.8 | **(1.85, 1135)** |
|  | Micro Acc.(%) | 6.95 | 4.4 | **(1.45, 12.44)** |
|  | Precision$_{wt.}$(%) | 12.65 | 11.3 | **(1.93, 23.38)\*** |
|  | Recall$_{wt.}$(%) | 6.95 | 4.4 | **(1.45, 12.44)** |
|  | F1$_{wt.}$(%) | 7.53 | 5.3 | **(0.93, 14.12)** |
|  | Iterations | 92.80 | 47.1 | **(34.32, 151.28)** |

Diff. Mean = Difference Mean; Diff. SD = Difference Standard Deviation;
CI = Confidence Interval; * = 90% CI

### 6.2 Results Analyses for Dataset2

Now, we analyse the results of Dataset2, as shown in Table 9, the three variant of the EM algorithm were ran using the corresponding optimal K, see the values of $K_2$ and $K'_2$, respectively in Fig. 3 and Fig. 4.

Again, the macro accuracy computed at the students' trace-level of the three model-based clustering methods is better than the Baseline model of Dataset2. These results suggest that model-based clustering has a potential to discover hidden patterns in diverse datasets and it is a good approach to construct mixture Markov models instead of a single model using the complete train data.

The predictive models for the EM and K-EM has achieved better precision results in comparison to the Baseline non-mixture model for Dataset2. However, in general, the Dataset2 seems to have more false-positives since the precision of all the predictive models is very low (ranging between 36% to 39% only). The emEM and K-EM has got better recall scores than the Baseline model; whereas the three model-based clustering methods has almost similar F1 score to that of the Baseline model in Dataset2.

If we compare the three model-based clustering methods, we can observe that the emEM and EM has achieved almost similar or better results in some

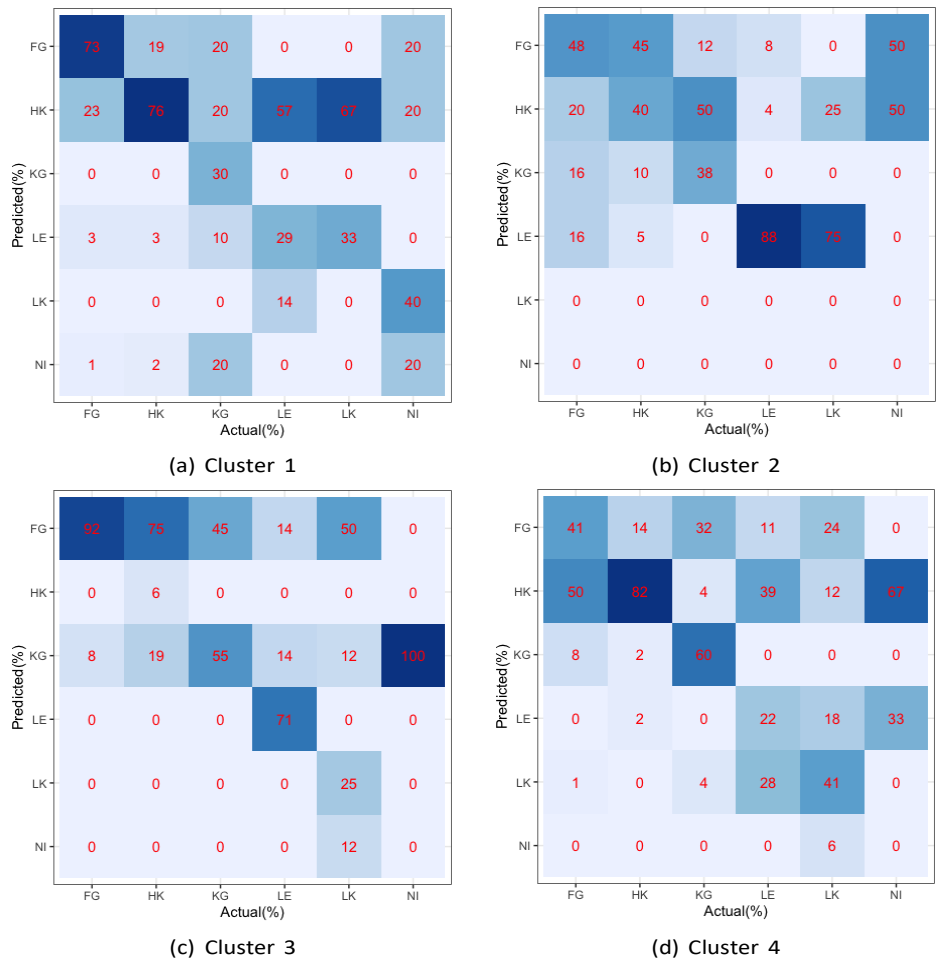

**Fig. 5:** Confusion matrices heatmap for the Dataset1 with $K'_1$=4 mixtures obtained using K-EM method

**Table 9:** Comparison of student-level 5-folds cross validation results for Dataset2 *(185 students; 348 traces; models constructed as described in Section 5; EM and emEM are run with $K_2$=3; K-EM with $K'_2$=2 mixtures; Baseline Model does not contain any cluster)*

|  | Baseline Model | | EM | | emEM | | K-EM | |
|---|---|---|---|---|---|---|---|---|
|  | Mean | SD | Mean | SD | Mean | SD | Mean | SD |
| Macro Acc.$_t$(%) | 47.71 | 4.2 | 52.09 | 3.9 | 53.08 | 4.3 | 51.05 | 2.6 |
| Micro Acc.(%) | 53.98 | 2.7 | 54.68 | 1.6 | 56.56 | 3.1 | 56.12 | 2.2 |
| Precision$_{wt.}$(%) | 36.88 | 2.7 | 38.9 | 3.3 | 36.7 | 5.7 | 38.31 | 5.7 |
| Recall$_{wt.}$(%) | 53.98 | 2.7 | 54.68 | 1.6 | 56.56 | 3.1 | 56.12 | 2.16 |
| F1$_{wt.}$(%) | 43.69 | 2.7 | 42.28 | 1.1 | 43.93 | 4 | 43 | 3.5 |
| Iterations | – | – | 38.2 | 22.2 | 48.2 | 27.7 | 28.8 | 36.4 |

SD = Standard Deviation

**Table 10:** Dataset2: 95% confidence interval for paired difference over 5-folds student-level cross-validation. A positive value in the "Diff. Mean" column shows that the K-EM method has better performance in difference means of the two methods; and a negative value meant otherwise (except for the "Iterations" metric where a low value represents a better performance).

|  | Metric | Diff. Mean | Diff. SD | 95% CI |
|---|---|---|---|---|
| K-EM/Baseline | Macro Acc.$_t$(%) | 3.35 | 3.3 | **(0.22, 6.47)*** |
|  | Micro Acc.(%) | 2.14 | 1.6 | **(0.15, 4.12)** |
|  | Precision$_{wt.}$(%) | 1.43 | 4 | (-3.58, 6.4) |
|  | Recall$_{wt.}$(%) | 2.14 | 1.6 | **(0.15, 4.12)** |
|  | F1$_{wt.}$(%) | -0.69 | 1.5 | (-2.55, 1.2) |
| K-EM/EM | Macro Acc.$_t$(%) | -1.03 | 3.4 | (-5.22, 3.15) |
|  | Micro Acc.(%) | 1.44 | 2.4 | (-1.48, 4.35) |
|  | Precision$_{wt.}$(%) | -0.58 | 5.8 | (-7.78, 6.62) |
|  | Recall$_{wt.}$(%) | 1.44 | 2.4 | (-1.48, 4.35) |
|  | F1$_{wt.}$(%) | 0.72 | 2.8 | (-2.77, 4.20) |
|  | Iterations | -9.40 | 38 | (-56.60, 37.80) |
| K-EM/emEM | Macro Acc.$_t$(%) | -2.02 | 2.9 | (-5.60, 1.56) |
|  | Micro Acc.(%) | -0.44 | 1.6 | (-2.44, 1.56) |
|  | Precision$_{wt.}$(%) | -1.34 | 3.3 | (-5.42, 2.74) |
|  | Recall$_{wt.}$(%) | -0.44 | 1.6 | (-2.44, 1.56) |
|  | F1$_{wt.}$(%) | -0.93 | 1.6 | (-3.22, 1.36) |
|  | Iterations | -19.40 | 25.2 | (-50.65, 11.85) |

Diff. Mean = Difference Mean; Diff. SD = Difference Standard Deviation;
CI = Confidence Interval; * = 90% CI

metrics than our proposed K-EM method. A probable justification that we consider for this performance degradation of the K-EM method in Dataset2 is that it has less number of mixtures in contrast to the EM and emEM methods, that is, $K'_2$=2 and $K_2$=3. In other words, one could have claimed that the high value of $K'_1$ in Dataset1 has got us improved results for the Dataset1 while it did not happen in the Dataset2 due to a low value of $K'_2$ in comparison to the EM and emEM methods. However, notice that the difference in mean and variance between the three methods in Dataset2 is less and therefore, it is not reasonable to conclude anything before doing further analyses.

Table 10 show our results for 95% confidence interval computed for the paired difference of each evaluation metric obtained using 5-folds student-level cross-validation for Dataset2. We reported the paired difference mean, paired difference standard deviation and 95% CIs for different methods in comparison to K-EM. The CIs ranges not involving the value of zero are shown in boldface, which indicates a significant performance difference between the two methods with 95% confidence.

We can see that the K-EM method has shown significant performance difference in comparison to the Baseline model of Dataset2, for Micro Acc. and Recall with 95% CI and with 90% CI for Macro Acc. $_t$. Whereas, the K-EM method achieved performance difference in Macro Acc.$_t$ with 90% CI. Like

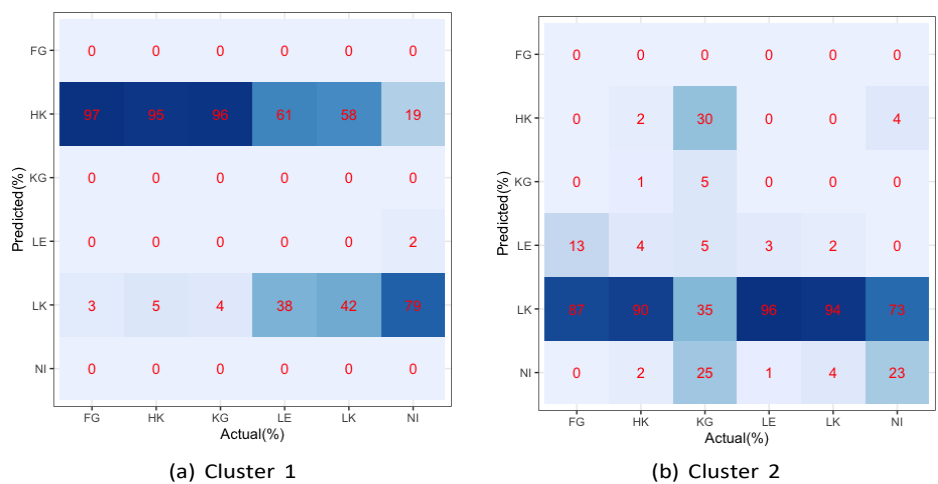

(a) Cluster 1      (b) Cluster 2

**Fig. 6:** Confusion matrices heatmap for the Dataset2 with $K'_2=2$ mixtures obtained using K-EM method

K–EM, the emEM has shown significant performance difference in comparison to the Baseline model of Dataset2 for Macro Acc.$_t$, Micro Acc. and Recall with 95% CI. Although, the EM method has shown significant performance difference for Macro Acc.$_t$ and Precision only. As we compared the K-EM with EM and emEM based on our results shown in Table 9, the latter two methods has shown somewhat better performance than our proposed approach. Though, as expected none of them has got significant performance difference in comparison to the K-EM method, despite having an increased value of $K$ as per the results given in Table 10.

Fig. 6 shows confusion matrices heatmap of the resultant $K$ mixtures for Dataset2 using the K-EM algorithm with $K'_2=2$. As we can see, the Markovian property based predictive model has performed even worse for Dataset2 which has longer trace lengths on average as compared to Dataset1 (see Table 3). The three least frequent behavioral patterns in the Dataset2 (i.e., KG, NI, and FG) has more false negatives. While, the two most frequent behavioral patterns (i.e., HK and LK) has more false positives and hence we got a very low mean weighted precision value for Dataset2, only 38.31% as shown in Table 9.

## 6.3 Discussion

Surprisingly, the two randomly initialized methods, EM and emEM, have shown no significant performance difference than the corresponding Baseline model of Dataset1, while both have shown at least some significant results in comparison to the Baseline model of Dataset2 (as discussed in Section 6.2). Whereas, our proposed K-EM method has shown significant performance difference for Macro Acc.$_t$, Micro Acc. and Recall$_t$, in comparison to the corre-sponding Baseline models of both datasets. The three model-based clustering methods have somewhat struggled in achieving better weighted Precision$_t$ values than the non-mixture Baseline models of both datasets. As shown by the confusion matrices previously, the problem lie due to high false positives for some most frequent behavioral patterns in $K$ resultant mixtures of imbalance datasets.

Therefore, in agreement to Cadez et al. (2003), we can conclude that mixture Markov models achieve better prediction accuracy results in comparison to a non-mixture first-order Markov chain. In other words, multiple predictive models can better capture varied behaviors depicted by students in comparison to a single predictive model constructed using the whole train data. Hence, it is useful to apply model-based clustering methods on students' problem-solving actions data which are usually diversify in nature.

In the above, we also compared the performance of the EM and emEM (which uses random initialization approach) with our proposed K-EM method that utilizes data-specific information for initialization from a preliminary K-means clustering algorithm. The three methods were ran on the two datasets using correspoinding optimal value of $K$, that is, for Dataset1: $K'_1$=4 for K-EM, $K_1$=2 for both EM and emEM); and for Dataset2: $K'_2$=2 for K-EM, $K_2$=3 for both EM and emEM). The K-EM has achieved overall better results in Dataset1 in comparison to the emEM method (see Table 8, with an exception of Precision$_{wt.}$). The K-EM method has also shown performance difference with 95% confidence interval in Macro Acc.$_{wt.}$, Micro Acc. and Recall$_{wt.}$ in contrast to the EM method. In case of Dataset2, our proposed K-EM method did not performed well neither worse than the EM and emEM methods despite having less number of mixtures.

Therefore, we can conclude that our proposed K-EM method has shown promising results in comparison to the two random initialization methods. However, the two datasets used in this work were relatively small in sizes (i.e. Dataset1 contains 197 traces of 92 students; Dataset2 contains 185 traces of 348 students) and suffered by class imbalance problem. Whereas, to prove a new algorithm/method, a more appropriate approach is to use a benchmark dataset(s) (Salzberg, 1997). In the future, we aim to perform comparative analyses of our proposed K-EM method with different approaches using large datasets.

Additionally, our detailed analyses of the confusion matrices for both datasets revealed that the Markovian property based predicted models for first-order Markov chain have struggled in producing correct predictions. This could be due to the fact that our datasets contained varied (dis)engagement behavioral patterns in the students' traces. Another probable reason could be the limitation of the classification scheme used in this work from Maqsood et al. (2019) to map students' problem-solving actions into six (dis)engagement behavioral patterns. We further discuss this limitation of the classification scheme in Section 9.

## 7 Visualizing and Interpreting Students' (Dis)Engagement Behavioral Patterns

In this section, we present visual representation and interpretation of the resultant mixture Markov models, obtained using the K-EM method with a corresponding optimal value of $K'$ (i.e., $K'_1=4$ for Dataset1; $K'_2=2$ for Dataset2) on complete datasets – Dataset1 with 197 traces and Dataset2 having 348 traces.

Fig. 7 and Fig. 8 contain Markov models[5] respectively for Dataset1 and Dataset2. The states of the Markov chains (shown by circles) represent six discrete (dis)engagement behavioral patterns. The size of each state is proportional to its support (or percentage of occurrence) in a specific cluster to show varied behaviors composed of some frequent and rare behavioral patterns as depicted by the students during assessment [6]. The thickness of an edge between two states is proportional to the transition probability between them (scaled by a constant factor). The transition probabilities greater than 32% are displayed only for legibility and to highlight prominent behavioral patterns and easy interpretation (also, we refer to the resultant $K$ mixtures as clusters in the following text). Moreover, in each Markov chain, the two behavioral patterns representing students' engagement (i.e., **FG** and **LE**) are shown by the states on the left; the two behavioral patterns representing their correct knowledge (i.e., **HK** and **LK**) are shown by the states in the middle; while the two behavioral patterns for students' disengagement (i.e., **KG** and NI) are shown by the states on the right in each figure. In the figures, we also made distinction between behavioral patterns related to the students' high or low confidence during assessment; that is, the upper-half of each figure shows behavioral patterns representing their low-confidence (such as: **LE, LK,** and NI), whereas, the lower-half of each figure shows behavioral patterns representing high confidence of the students during assessment (such as: FG, HK, and KG).

In the following sub-sections, we interpret the students' (dis)engagement behavioral patterns as depicted from their logged interactions in each dataset. However, we only consider the frequent behavioral patterns (based on the sizes of their corresponding states in a Markov chain for making correct interpretations.

---

[5] All plots were drawn using r-igraph: https://igraph.org/r/.

[6] Furthermore, states are filled with different colors to highlight their meanings. For example engagement behavior reflected with either confidence level is represented by two states, FG and LE, which are given the same color (yellow) in the images. Similarly, states representing disengagement behaviors: KG and NI, are shaded with the same color (blue). High knowledge (HK) and less knowledge (LK) states are differentiated with gray and white colors, respectively; *see colored pictures in online PDF version.*

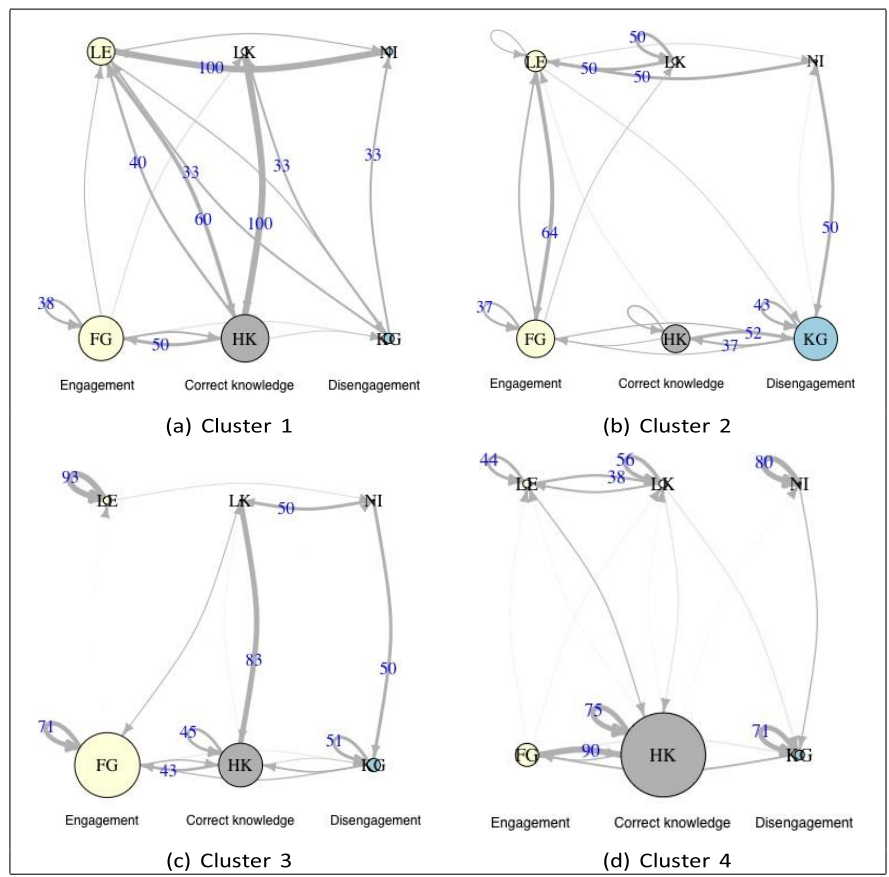

**Fig. 7:** Four resultant Markov chains for Dataset1 (using complete dataset): (a) Cluster 1: *8 traces* ; (b) Cluster 2: *27 traces* ; (c) Cluster 3: *101 traces* ; (d) Cluster 4: *61 traces* ; The size of each state is proportional to its percentage in the respective cluster and thickness of each edge is proportional to the transitional probability between respective states (scaled by a constant factor).

### 7.1 Interpretation of the Students' (Dis)Engagement Behavioral Patterns in Dataset1

In Fig. 7(a), Cluster 1 represents the smallest group of students' traces – only 4% traces of the complete Dataset1. The frequent behavioral patterns are high knowledge (HK) and fill-knowledge gap (FG) which shows that the traces belong to high confident students who depicted positive engagement during assessment – as shown by more incoming edges to the HK state and transition probabilities between HK and FG. This is the only cluster which also contain some representation of the learn (LE) behavioral pattern that shows positive engagement of low confident students.

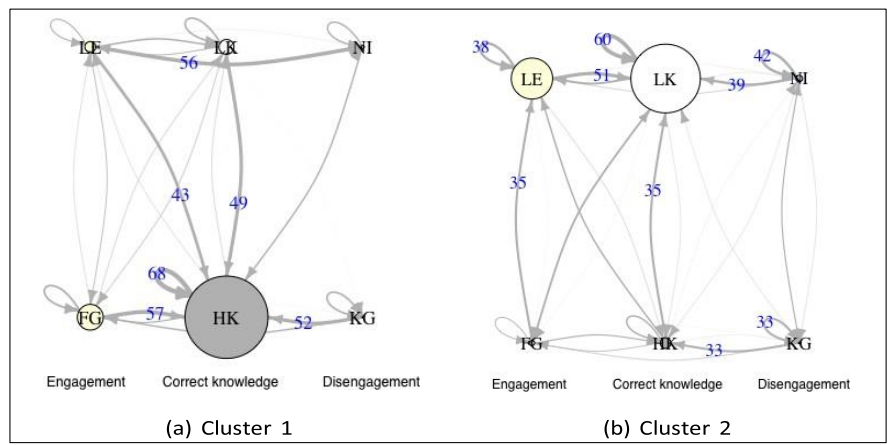

**Fig. 8:** Two obtained Markov chains for Dataset2 using 85% train data: (a) Cluster 1: *176 traces* ; (b) Cluster 2: *172 traces* ; The size of each state is proportional to its percentage in the respective cluster and thickness of each edge is proportional to the transitional probability between respective states (scaled by a constant factor).

Cluster 2, shown in Fig. 7(b), represents the second least similar group of students' traces (i.e., 14% of the complete Dataset1). The most frequent behavioral patterns in this cluster are knowledge gap (KG) and fill-knowledge gap (FG), representing respectively the disengagement and engagement of high confident students. The KG state has more incoming edges (one is from the HK state with 52% transition probability) including a self-loop of 43% transition probability. This shows that the students having high confidence in wrong answers did not request the available (task-level) feedback which could have helped them in learning from their mistakes and performing better in the subsequent questions. Hence, it reflects their disengagement during the assessment. Another observable activity in this cluster is fill-knowledge gap (FG) with an incoming edge from LE state with 64% transition probability. Since, both LE and FG states represent the students' engagement with low and high confidence, respectively; we interpret this behavior as change in one's confidence level from low to high.

Cluster 3, shown in Fig. 7(c), is the largest subgroup of traces in the Dataset1 (i.e., 51% traces) which depicts positive engagement of the students during assessment. The traces contain the fill-knowledge gap (FG) behavioral pattern as the most dominant behavior followed by the high knowledge (HK). The FG behavioral pattern reflect that the students attempt to fill their knowledge gap(s) through detailed (task-level) feedback (mainly) for wrong answers (Maqsood and Ceravolo, 2019). A high transition probability of a self-loop on FG state (i.e., 71%) shows that students in case of wrong response(s) majorly focused on learning from the task-level feedback available for each submitted problem. Also, in cases when the students show high knowledge (HK), they moved to the fill-knowledge gap (FG) behavioral pattern for incorrect answers. Finally, edges from states in the upper-half to their corresponding states in the lower-half show a change in the students' confidence level from low to high in a respective knowledge state – for example: an edge with 83% transition probability from LK to HK and the one with 50% transition probability from NI to KG. The first one, that is, an edge from LK to HK is indeed desirable, that is, an under-confident student who might be answering the questions correctly with a low confidence, should improve his/her confidence-level in the subject domain over time.

The second largest subgroup of the students' traces found in Dataset1 is shown by Cluster 4 in Fig. 7(d), which comprises of 31% of the total traces in the dataset. Since, the most frequent behavioral pattern occurred in this group of traces is high knowledge (HK), we can say that these traces belong to the students having high knowledge in the subject domain who gave more correct answers with high confidence (see the HK state's size and a high probability self-loop, i.e., 75%).

### 7.2 Interpretation of the Students' (Dis)Engagement Behavioral Patterns in Dataset2

Cluster 1, shown in Fig. 8(a), represents 51% of the total students' traces in the Dataset2. These traces mainly reflect high knowledge of the respective students (see the size of the HK state). The HK state has many incoming edges and a a high transition probability self-loop (i.e., 75%). Note that the students' behaviors reflected by this Markov chain are quite similar to the second largest subgroup of traces found in the Dataset1 (see Cluster 4 in Fig. 7(d)).

Cluster 2 , shown in Fig. 8(b), is comprised of 49% of the students' traces in the Dataset2. The students' traces in this subgroup reflect correct knowledge and positive engagement of the low confident students – as shown by the less knowledge (LK) and learn (LE) behavioral pattern states, respectively. The LK state also has a self-loop with 60% transition probability which shows that the respective students correctly answered the subsequent questions with low confidence. We can assume that those students have doubts attained correct knowledge of the subject domain but they have doubts about it (Gardner-Medwin and Gahan, 2003). Learning (LE) is the second most frequent behavioral pattern observed in this cluster which shows engaged behavior of some low confident students during assessment.

### 7.3 Summary

In summary, visualization of the resultant mixture Markov models provides substantial insights about the students' (dis)engagement behaviors in both datasets. Through these Markov chains, a class teacher can better understand strengths and weaknesses of his/her students by visualizing different subgroups of distinct behavioral patterns. For example, this could be a point of concern for a class teacher to further investigate the potential reason(s) for the high ratio of traces with low confidence observed in the Cluster 2 of Dataset2 (i.e., 49% of the total dataset). In our opinion, it could be either due to (high) difficulty level of the posed questions or the perceived toughness of the course by the respective students, which made them felt low confident about their (correct) knowledge. Similarly, some students having high confidence in wrong responses depicted disengaged behavior during the assessment (see Fig. 7(b) – Cluster 2 of Dataset1). The intervention of a class teacher is required in this case to understand why those respective students did not requested for a task-level feedback for the questions answered incorrectly during computer-based assessment.

## 8 Related Works

### 8.1 Measuring Student Engagement

There are several methods used in the existing literature for data collection and estimating students' engagement behavior. For example, (Chapman, 2003) reported a number of alternative methods used by the researchers, including students' self-report engagement level (through questionnaires), checklists and rating scales - done by the teachers, direct observations of students in a class, (students') work sample analyses (e.g. project, portfolio, etc.), and, case studies. As mentioned earlier, our focus is on analyzing students' interactions data recorded by a computer-based assessment system. Therefore, in the following, we discuss attributes and methods used to measure student engagement by related works only which have taken students' logged data as an input.

Hershkovitz and Nachmias (2009) referred to engagement as an attribute of motivation during learning and used Hierarchical clustering algorithm to identify the best attributes that mapped on existing theories of motivation. They identified the following two variables to determine student engagement: *time on task percentage* and *average session duration*. Cocea and Weibelzahl (2009) also linked engagement with students' motivation in a subject or domain and estimated it using: *frequency* and *effort (or time) spent* on both reading pages and quizzes attempted by the students as they interacted with three different learning environments. Students' sessions were labeled as *engaged* or *disengaged* by human experts based on a set of rules defined earlier from manual analysis of the data (Cocea and Weibelzahl, 2007). Eight data mining techniques were then used to construct a prediction model for student (dis)engagement, for example, Bayesian nets, Logistic regression, Decision tree, etc. Their supervised approach relied on pre-analysis of the data performed by human experts to identify a suitable length of traces which is data-dependent. Hence, the re-usability of the implemented method is reduced extensively. Whereas, we adopted an unsupervised approach using a probabilistic model that takes care of traces of different lengths.

Beal et al. (2006) adopted the notion of students' active participation in a current task and classify students' problem-solving activities into five different levels of engagement using: *response correctness*, *time spent per problem* and *help usage*. Hierarchical clustering was applied to proportion scores of these patterns to analyze students' use of an intelligent tutoring system (ITS). Another experimental study presented in (Brown and Howard, 2014) uses on-/off- task notations to refer to engaged and disengaged behaviors, respectively. Specifically, they used *response correctness*, *time on task* and *triggered events (i.e., keyboard strokes and/or mouse movements)*; attributes to label students' actions as engaged or disengaged. Engagement is considered as one of the affective states in (Pardos et al., 2014) which is determined using *number of correct answers*, *proportion of actions in a time frame*; *number of reattempts*, *hints requested* and *fail on first attempt*. Human experts' (in field) observations were synchronized with student logged data to define a mapping between recorded interactions and various affective and behavioral states observed by the experts. Eight classification methods including Decision trees, Naive Bayes, Step regression and others were used to build a model for automatic detection for each effective state separately.

The literature review shows the potential of students' logged interactions to determine their level of involvement in the learning process. However, the classification of students' problem-solving activities into engagement/disengagement behaviors depends on the problem domain and collected data attributes. As mentioned earlier, we used a classification scheme defined in (Maqsood et al., 2019) for mapping students' problem-solving activities into six behavioral patterns reflecting their engagement and disengagement during confidence-based assessment. Our work is distinguished from prior works as we have analyzed sequential traces of students' interactions to understand their progression from one behavioral state to another using a more sophisticated probabilistic model.

8.2 Modeling and Predicting Humans' Behaviors using Probabilistic Methods

Although several techniques have been presented in the literature to extract meaningful information from students problem-solving traces recorded by computer-based learning environments, for example: clustering (Beal et al., 2006; Hershkovitz and Nachmias, 2009; Köck and Paramythis, 2011; Boroujeni and Dillenbourg, 2018), classification (Cocea and Weibelzahl, 2009, 2011; Pardos et al., 2014; Maqsood et al., 2019), evolutionary method (Romero et al., 2004), Bayesian network (Muldner et al., 2011), deep learning and other machine learning techniques (Botelho et al., 2019), etc. Our focus in on the family of probabilistic approaches used to model and/or predict human behavior. In this section, we discuss some applications of different methods specifically including Markov chain, hidden Markov model and mixture of Markov chains.

Authors in (Taraghi et al., 2015) modeled students' question answering patterns (i.e. right or wrong answer) using second-order Markov chains to construct their profiles. Another application of Markov chains to capture and predict users' behaviors is given in (Khalil et al., 2007), where each trace contains a user's navigational pattern on a website. A simple K-means algorithm is used to group users having similar web navigation behaviors. Each cluster is then represented by a Markov chain and a user's future behavior is predicted accordingly. Their work is limited as it restricts a user's behavior to be represented by only one Markov chain. Whereas, our approach of clustering similar Login-Logout sessions using mixture Markov chains allows the flexibility of capturing change in a student's behavior from one session to another. Furthermore, model-based clustering is a more sophisticated method to group traces of different lengths in contrast to distance-based clustering approaches like K-means and Hierarchical clustering algorithms (Cadez et al., 2003) used in some prior works, e.g., (Khalil et al., 2007; Taraghi et al., 2015).

Simple Markov chains are restricted to observable data only, whereas, sometimes it is important to identify underlying hidden information to represent the internal cognitive behaviors of the users. Hidden Markov Model (HMM) is another very popular probabilistic approach amongst researchers to analyze and model humans' behaviors, where the hidden or latent states overcome the pre-mentioned limitation of Markov chains. For example, (Beal et al., 2007) captured students' problem-solving behaviors using HMM where latent states reflect their different levels of engagement (i.e. low, medium, high) with an ITS. Also, in (Fok et al., 2005) a classification model is developed using a hidden Markov model to characterize students showing different content access preferences while interacting with an e-learning system.

Bouchet et al. (2013) used the Expectation-Maximization (EM) algorithm to cluster students' profiles participating in a self-regulated learning environment. Although resulted clusters reveal distinct behavioral patterns of the students, sequential ordering of the activities is not considered by the authors which may have offered useful insights to further distinguish between students and improve system adaptation. Cadez et al. (2003) also utilized model-based clustering to analyze web navigation patterns of website users where each trace contains the sequential ordering of web pages accessed by a user. Their approach is quite related to that of ours in a way that they also used mixture of first-order Markov chains to model and analyze sequential categorical data representing users' dynamic behaviors. However, our method is a modification to the original EM algorithm which improves the prediction accuracy for each resultant cluster.

Recent work on understanding students' procrastination behavior (Park et al., 2018) has utilized model-based clustering where each mixture component follows a Poisson distribution to show students' activities in an online course. Hansen et al. (2017) also used mixture of Markov chains to model dynamic behaviors of the students captured by an e-learning system. Their proposed method estimates mixture components (i.e. first-order Markov chains) using a modified K-means clustering algorithm. The authors made a similar assumption that students' behaviors may change over time and thus performed activity sequences analyses at the session level, which associates mul-tiple Markov chains with an individual student representing his/her different problem-solving sessions. Despite having some similarities, our approach is an extension to the standard EM algorithm which is more accurate for estimating the likelihood of related sequential traces and generates (a mixture of) Markov chains with better prediction accuracy.

## 9 Summary, Conclusion and Future Work

This research work aimed to analyze, model and predict students' (dis)engagement behaviors in confidence-based assessment. The two datasets used in this work came from two experimental studies conducted with undergraduate students from Pakistan and Italy. The two studies were conducted using computer-based assessment tools which logged student-system interactions as the students performed any activity during assessment. Using the classification scheme introduced in Maqsood et al. (2019), students' problem-solving actions were then classified into six (dis)engagement behavioral patterns, that is, the set $P$={HK, LK, FG, LE, KG, NI}. The previously proposed scheme in Maqsood et al. (2019) considers a student's three problem-solving actions to map it to a corresponding (dis)engagement behavioral pattern, including: a student's response correctness (i.e., a question answered correctly or not), his/her associated confidence-level as high or low with a submitted response, and whether a student has requested task-level feedback subsequently for the question answered most recently. Thus, the two datasets used in this work contain students' traces of different lengths wherein each event represents a corresponding engaged or disengaged behavioral pattern from the set $P$.

In this work, we employed model-based clustering to find subgroups of students' sequential traces; wherein each trace contains a sequence of varied (dis)engagement behavioral patterns depicted by a student during computer-based assessment (see Table 4 for sample data). The Expectation-Maximization (EM) algorithm used for constructing mixture Markov models struggle for finding 'global' maxima and hence the initialization method can play an important role in finding a best solution (Michael and Melnykov, 2016; Hu, 2015). Thus, in this work, we proposed a new initialization method called "K-EM" that uses the results of a preliminary K-means clustering algorithm to initialize the EM algorithm for multivariate categorical data (as explained in Section 3) .

In Section 6, we report our results of the experiments performed using the K-EM method, the two existing EM algorithm variants namely, the original EM and emEM, and non-mixture baseline models for the two datasets. The $K$ mixture Markov models are constructed for both datasets using 5-folds student-level cross-validation using the EM, emEM and K-EM methods. For each resultant mixture (or cluster), we then constructed a corresponding first-order Markov chain – which is used by the predictive model to predict a student's future behavioral pattern for each *test data* trace (see our methodology in Fig. 1).

Our results showed that the K-EM has achieved significantly better prediction accuracy (both micro and macro) and recall than the corresponding non-mixture baseline models for both datasets (as shown in Table 8 and Table 10. Also, for Dataset1, our proposed K-EM method has shown overall significant performance improvement than the emEM except that the emEM converges faster. The K-EM method has also shown significant performance difference with 95% confidence interval for Macro Acc.$_{wt.}$, Micro Acc., Recall$_{wt.}$ and F1$_{wt.}$ in contrast to the EM method, see Table 8. While for Dataset2, the K-EM method has achieved slightly poor or sometimes almost equal results in comparison to the EM and emEM methods. However, there is no evidence of significant mean paired difference between the results of the three methods on different performance evaluation metrics, see Table 10. Thus, we conclude that our proposed K-EM method has shown promising results in comparison to the two randomly initialized methods. However, the two datasets used in this work were relatively small in sizes (i.e. Dataset1 contains 197 traces of 92 students; Dataset2 contains 185 traces of 348 students) and both datasets have class imbalance problem. To conclude, our proposed initialization method for the Expectation-Maximization has captured the students' behavioral dynamics at a low interaction level. In other words, we better know the engagement or involvement level of a student (using a confidence-based assessment perspective) that track the confidence and engagement trajectories followed by students. We are optimistic that our methodology will have a positive influence on adaptive algorithms as our approach demonstrates a way of getting explainable results using data mining techniques.

The number of iterations required for model training by the three methods seems to correlate with the number of mixtures ($K$), where an increase in the value of $K$ will require more iterations in the model training phase on a dataset. The performance difference between the convergence rates of the three methods is not significant except for one case (i.e., emEM converges faster than K-EM for Datset1). We, therefore suggest to perform further analyses for different values of $K$ using different datasets.

A limitation of our work lies in our assumption that a student's future (dis)engagement behavioral pattern is only dependent on his/her most recent behavioral pattern during assessment. Given the heterogeneous nature of students' (dis)engagement behavioral patterns (as shown in Fig. 7 and 8), increasing the prediction performance for each obtained cluster is also a future challenge. A naive approach to further improve the prediction accuracy is to use a higher order Markov chains. But, an increase in the accuracy would come with a cost of an increase in time and space complexity which is not favorable especially if the developed model is to be implemented in an online setting (e.g., an adaptive system). Another possibility could be to utilize and evaluate other machine learning algorithms for building a predictor model for multivariate time-series data.

Additionally, the classification scheme used in this work from (Maqsood et al., 2019), for mapping the students' logged problem-solving actions into six (dis)engagement behavioral patterns is restricted. The classification scheme defines six (dis)engagement behavioral patterns namely: high knowledge (HK), less knowledge (LK), fill-knowledge gap (FG), knowledge gap (KG), learn (LE), and, not interested (NI); based the following three problem-solving actions: a student's response correctness (correct or incorrect), confidence level (high or low) specified for each submitted answer, and, a followed feedback-seeking activity (whether a student requested task-level feedback or not for an answered question). The *feedback-seeking activity time* was not considered in classification scheme due to lack of any evidence for its significant correlation with students' confidence-level in (Maqsood and Ceravolo, 2019). However, a minimal threshold could be defined on *feedback-seeking activity time* before classifying a student's problem-solving actions into engaged or disengaged behavioral pattern. Since, there is a possibility that a student just clicked on the task-level feedback page for curiosity (or let's say by mistake), or do not spend sufficient time to read and process the presented information, e.g., let's say below 10 seconds.

Finally, visualization of the resultant mixture Markov models reveal very useful insights for class teachers about students' (dis)engagement behavioral patterns, as discussed in Section 7. Implementation of these plots in an online assessment tool would provide easy access to various analytic to a class teacher after each computer-based assessment session. A teacher can identify strengths and weaknesses of students and may modify his/her teaching strategy accordingly. Also, the developed method can be implemented in an adaptive system that can identify students with undesirable behavior and offers personalized feedback to diverse groups of students. However, it may be difficult to provide any assistance in some cases, e.g. Cluster 2 of Dataset1 (see Fig. 7(b)). Here, we also highlight that the two larger subgroups of both datasets (i.e. Cluster 4 and Cluster 1 respectively of Dataset1 and Dataset2) reveal very similar behaviors of the students belonging to different populations. This is very promising for constructing a mixture of Markov models representing the most common behaviors of students through different mixture components, which can be identified by the domain expert(s). And, each new student can then be assigned to a suitable mixture component after collecting his/her problem-solving actions. Evaluating the prediction accuracy and testing this model on different populations is also a point of investigation for future work.

In the end, visualization of the resultant mixtures for both datasets (shown in Fig. 7 and 8) reveal that the students depicted varied (dis)engagement behavioral patterns in different Login-Logout sessions (or traces). Thus, in agreement to (Hansen et al., 2017), we conclude that it is advantageous to analyze students' interactions at the lowest representation level, i.e. activities contained in Login-Logout sessions. The mixture Markov models yielded through model-based clustering is a useful mechanism to capture students' diverse behaviors. But, it does not tell us about a student's transition(s) from one mixture component to another, which will allow us to construct a student's "personalized behavioral profile". Hidden Markov Models (HMM) has the advantage of having hidden states that are related by a Markov process and not just individual mixture components (Rabiner and Juang, 1986). Baulm-Weltch algorithm (which is based on the EM algorithm) besides inferring the model parameters, also infers transition probabilities between different hidden states. Hence, the flexibility of constructing users' profiles through HMM is favorable for researchers aiming to make practical use of their constructed model(s) in an adaptive learning system instead of just performing post-experiment(s) data analyses. However, training HMM models is computationally expensive in comparison to mixture Markov models. In future work, we intend to construct students' personalized behavioral profiles using mixture Markov models to represent their level of knowledge and (dis)engagement behaviors across different Login-Logout sessions. This will led us to comprehend a student's overall behavior from different Login-Logout sessions and we can better identify strengths and weaknesses of a student at a high level. For example, students who answer questions mostly with high or low confidence can be identified as having a specific confidence level as a personality trait or in the subject domain, instead of specifying his/her confidence accurately for each answered question. A student's personalized behavioral profiles will also allow us to understand any drift or change in his/her (dis)engagement behavioral patterns across different sessions. Hence, these profiles can also be used to provide personalized feedback at a high level accordingly.

**Acknowledgements** We would like to sincerely thank the three anonymous reviewers for providing their detailed and thoughtful comments, which has helped us to significantly improve this revised manuscript.

## Declarations

### Funding

### Conflict of interest

### Availability of data and materials